\documentstyle[11pt,vfigs]{article}

\newcommand{\be}{\begin{equation}}
\newcommand{\ee}{\end{equation}}
\newcommand{\bea}{\begin{eqnarray}}
\newcommand{\eea}{\end{eqnarray}}
\newcommand{\psibarpsi}{\left\langle\overline{\psi}\psi\right\rangle}

\begin{document}

\title{\bf The Role of the Polyakov loop \\ in Finite Density QCD}

\author{Ph. de Forcrand and V. Laliena \\
ETH, CH-8092 Z{\"u}rich, Switzerland}

\date{July 11, 1999} 
\maketitle

\begin{abstract}
We study the behavior of the fermion determinant at finite temperature and
chemical potential, as a function of the Polyakov loop. The phase of the
determinant is correlated with the imaginary part of the Polyakov
loop. This correlation and its consequences are considered in static QCD,
in a toy model of free quarks in a constant $A_0$ background, and in
simulations constraining the imaginary part of the Polyakov loop to zero.
\end{abstract}


\section{Introduction} 

The experimental study of QCD at finite density and temperature, and the
possible observation of the quark-gluon plasma are the focus of activity
of an increasing fraction of the physics community. A parallel study by
computer simulations would be highly desirable. However, after integration 
over the fermions, the integrand $\rho(U)$ in the finite density 
partition function is complex. The measure used in Monte Carlo simulations
is $\langle .. \rangle_{MC} = d[U]~|\rho(U)|$, and observables are computed as
$\langle {\cal O} \rangle = \langle {\cal O} \rho(U)/|\rho(U)| \rangle_{MC}
/ \langle \rho(U)/|\rho(U)| \rangle_{MC}$.
The denominator is exponentially small in the 4-volume of the lattice,
and measuring it with a given accuracy requires an exponentially growing
amount of statistics. This is the so-called sign problem. Over almost
twenty years, progress in controlling the sign problem has been negligible.

For this reason one has turned to approximations to QCD: the strong coupling
limit, and more recently the static limit of QCD have attracted
a lot of interest \cite{Blum,Karsch}. Static QCD is still affected by the
sign problem, which limits investigations to small volumes, far from the
continuum limit. It would be very desirable to extend the domain of study
of static QCD to more realistic regions of parameters, and to go beyond 
the static approximation and allow for at least some spatial hopping of the
quarks. Our paper proposes and tests a method to do this.

We have analyzed static QCD results of \cite{Blum} and found a strong correlation
between the phase of the determinant $\phi(\det)$ and the imaginary part $P_i$
of the Polyakov loop $P$. We study this correlation in Section I, and show that
it is to be expected for heavy quarks.

This provides the motivation for a simple toy model, studied in Section II,  
where all degrees of freedom are suppressed except the Polyakov loop: 
free quarks in a gauge field background with constant temporal component 
$A_0$ and vanishing spatial components.
This model can be solved easily in Fourier space. It shows several features of
QCD. In particular, it exhibits the ordering effect of the chemical potential 
$\mu$,
similar to that of the temperature $T$. Under simple assumptions, a phase 
diagram $\mu_c(T)$ can be obtained at small $\mu$. A mechanism for a 
transition at low temperature is proposed. 

Finally, the correlation between the phase of the determinant and the imaginary
part of the Polyakov loop can be put to profit by performing constrained 
simulations, where $P_i = 0$. The sign problem is reduced. The validity of 
imposing such a constraint, and the benefits this constraint provides, are 
reviewed in Section III.


\begin{figure}[!b]
\begin{center}
\begin{minipage}[h]{6 cm}
\vspace*{-1.0 cm}
\epsfig{bbllx=4.0truecm,bblly=9.3truecm,bburx=16.5truecm,bbury=21.95truecm,
file=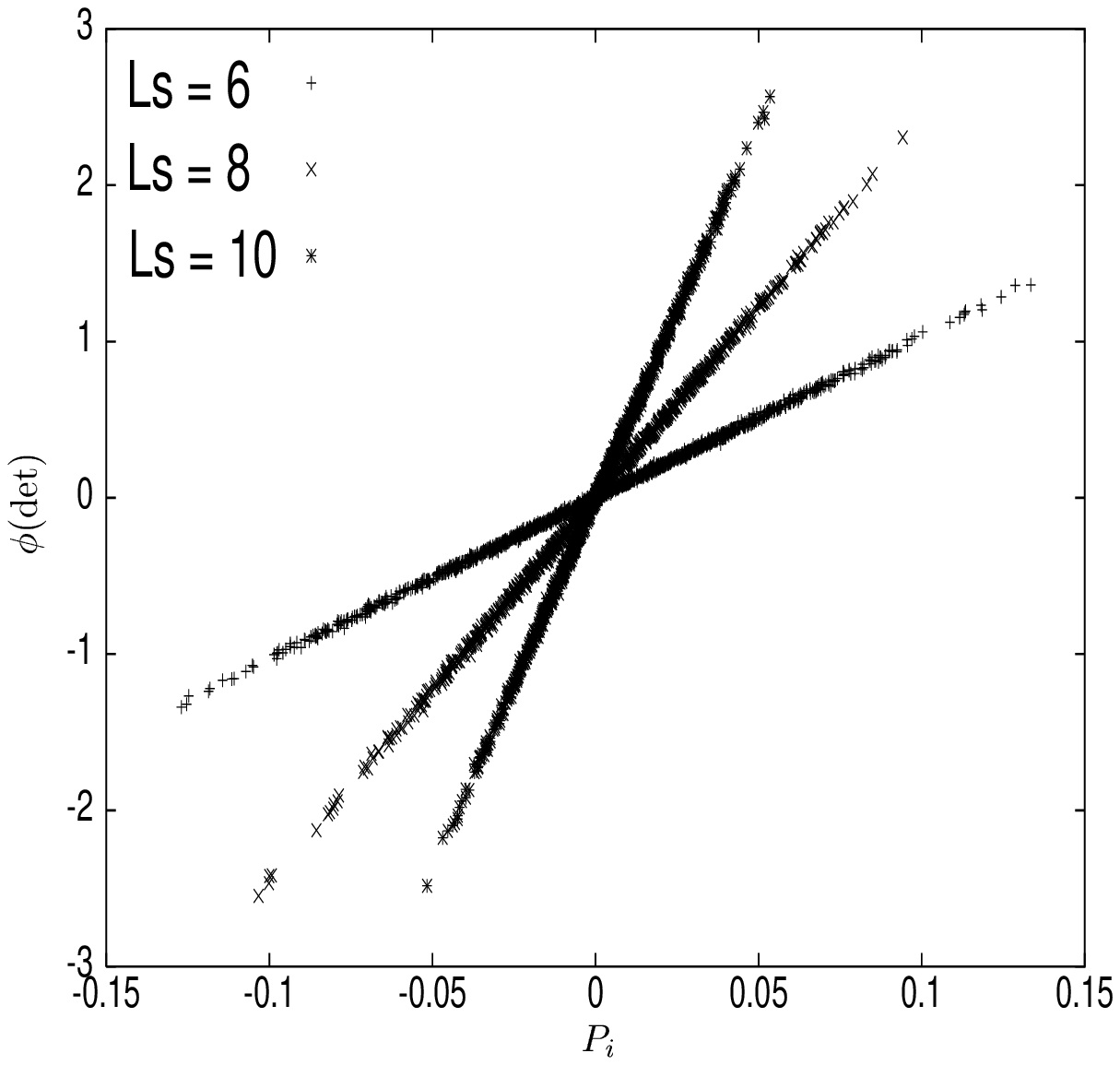,height=5.5truecm,width=5.5truecm,clip=}
\end{minipage}
\begin{minipage}[h]{6 cm}
\vspace*{-1.0 cm}
\epsfig{bbllx=4.0truecm,bblly=8.8truecm,bburx=16.5truecm,bbury=21.95truecm,
file=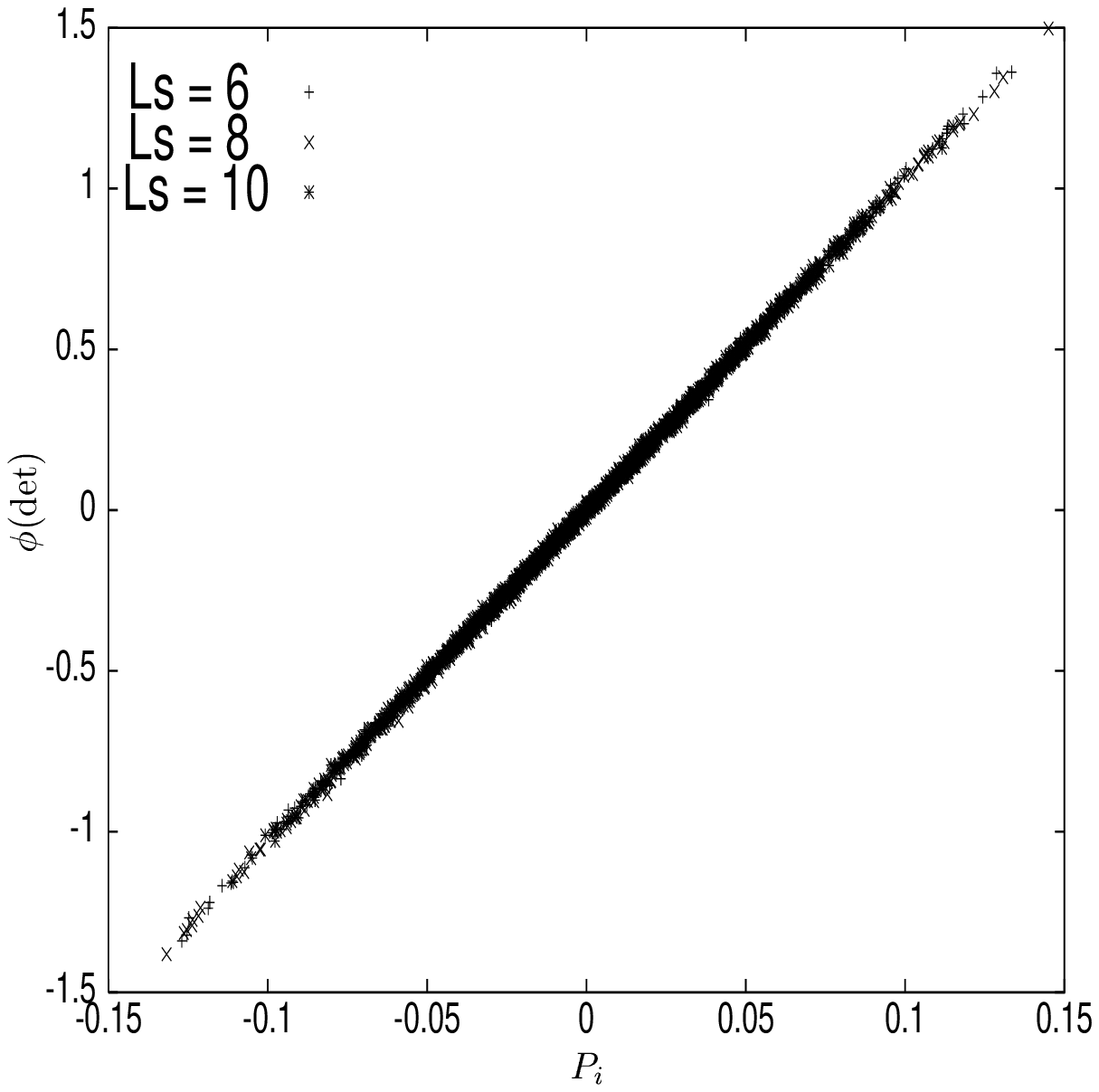,height=5.5truecm,width=5.5truecm,clip=}
\end{minipage}
\end{center}
\vspace*{-0.75cm}
\caption{\protect{
Correlation between $\phi(\det)$ and $P_i$ in static QCD, 
on a lattice $L_s^3\times 2$ ($L_s=6,8,10, C=1/20, \beta=5.0$).
On the right, $P_i$ has been rescaled by $\sqrt{L_s^3}$ and $\phi(\det)$
by $1/\sqrt{L_s^3}$. 
The data are from Ref.[1].
}}
\end{figure}


\section{Correlation between $P_i$ and $\phi(\det)$}

We have reanalyzed the static QCD data of \cite{Blum}.
Recall that static QCD is obtained by taking the quark mass $m_q$ and
the chemical potential simultaneously to infinity, such that the quark
density is finite. This is achieved by keeping constant the parameter
$C^{-1} \equiv (e^{\mu}/2m_q)^{N_t}$, where $N_t$ is the temporal
extent of the lattice.
The quarks are static, but the fermionic determinant is still
sensitive to changes in the gauge field and the quark density.
Fig.1a shows almost perfect correlation between the phase of the determinant 
and the imaginary part of the Polyakov loop, for 3 spatial volumes $V$.
Furthermore, as $V$ is increased, one observes that the fluctuations in $P_i$ 
vary like $1/\sqrt{V}$, as any extensive variable, and that fluctuations 
in $\phi(\det)$ increase like $\sqrt{V}$. Under this rescaling, the results
for the 3 different volumes become indistinguishable (Fig.1b).

\begin{figure}[t]
\begin{center}
\epsfig{bbllx=4.0truecm,bblly=9.5truecm,bburx=16.5truecm,bbury=21.5truecm,
file=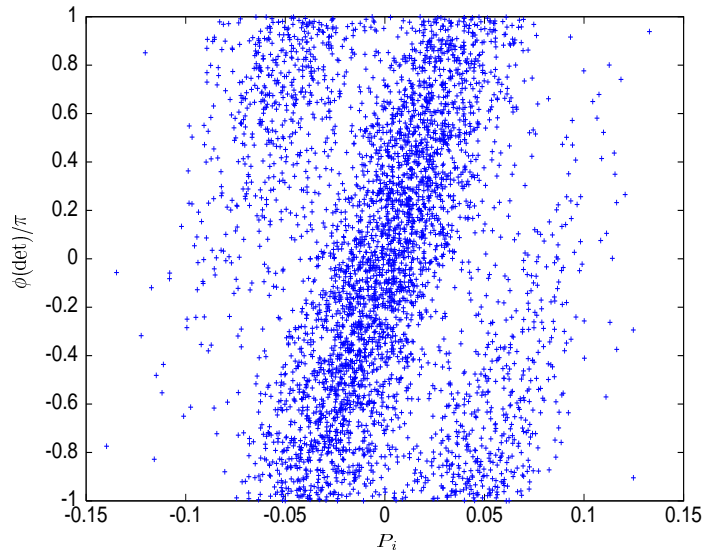,height=7.5truecm,width=9.5truecm,clip=}
\end{center}
\vspace*{-0.55 cm}
\caption{Same as Fig. 1 with $C=1/0.6$ and $\beta=5.0$ on a $6^3\times4$ 
lattice.}
\end{figure}

As the chemical potential (equivalently $C$ in static QCD) 
increases, or as the temperature decreases, fluctuations in the phase increase,
and the correlation becomes harder to unravel because of the compact nature
of the phase. Fig.2 shows the worst case studied by Blum et al. \cite{Blum}.
If one defines $\phi(\det)$ as $\sum_i \phi(\lambda_i)$, with
$\phi(\lambda_i) \in ]-\pi,+\pi]$, where $\lambda_i$ are the eigenvalues of 
the Dirac matrix, the correlation remains significant.

This correlation should be expected. In the loop expansion of the fermion
determinant, $\log (\det({\bf 1} - \kappa M) \equiv - S_{eff} = 
- \sum_l \kappa^l/l ~{\rm Tr} M^l$, where $\kappa \sim 1/m_q$, the first loop
affected by the chemical potential $\mu$ is the (timelike) Polyakov loop.
The corresponding contribution to $S_{eff}$ is
\bea
2 (2 \kappa)^{N_t} \sum_x (\cosh(N_t \mu) P_r(x) + i~\sinh(N_t \mu) P_i(x)) \\
{\rm i.e.} ~~~~~\phi(\det) = 2 (2 \kappa)^{N_t} V ~\sinh(N_t \mu) P_i
\, ,
\eea
where the summation $\sum_x$ is over all spatial coordinates $x$.
Higher-order terms spoil this perfect correlation. They become more
important as $N_t$ increases or $m_q$ decreases.
In static QCD, there is no spatial hopping of the quarks, so that the
only additional terms are powers of the Polyakov loop at each site, which
are suppressed by powers of $2 (2 \kappa)^{N_t} ~\sinh(N_t \mu)$.

\section{Toy model: free quarks in a constant $A_0$ background}

We have just seen that static QCD is in fact a Polyakov loop model, 
where the phase of the determinant is almost equal to the imaginary part
of the Polyakov loop. But even in static QCD, calculations are still
severely limited by the sign problem. Monte Carlo integration performs
an averaging over the gauge fluctuations and the fluctuations of the
Polyakov loop $P(x)$ at each spatial point $x$. We can simplify this
model further by keeping only the zero-mode of such fluctuations and
neglecting, for now, the other fluctuations of the gauge field.
We end up with a single $SU(3)$ degree of freedom, the Polyakov loop $P$.
Our model consists of free quarks in a gauge field background with
zero spatial components and constant temporal component $A_0$.
Hence, $P = \exp(i N_t a A_0)$, where $a$ is the lattice spacing. Gauge
invariance implies that the fermion determinant depends only on
${\rm Tr} P = P_r + i P_i$. It can be calculated by Fourier transform on the
lattice, using staggered quarks. As this work was in progress, the preprint 
\cite{Pisarski} appeared, where the same calculation of the determinant 
is performed in the continuum. This determinant represents the fermion 
contribution to the one-loop effective potential of the Polyakov
loop. It should become a better approximation 
as the gauge fluctuations are suppressed, i.e. at high temperature.
The behavior of the determinant as a function of $(P_r,P_i)$ is 
summarized in Figs.3 to 5 for $m_q=0$. Increasing the quark mass does not
change the qualitative behavior, but only reduces the amount of variation
of the determinant.

\begin{figure}[!h]
\begin{center}
\epsfig{bbllx=5.5truecm,bblly=2.85truecm,bburx=16.5truecm,bbury=24.5truecm,
file=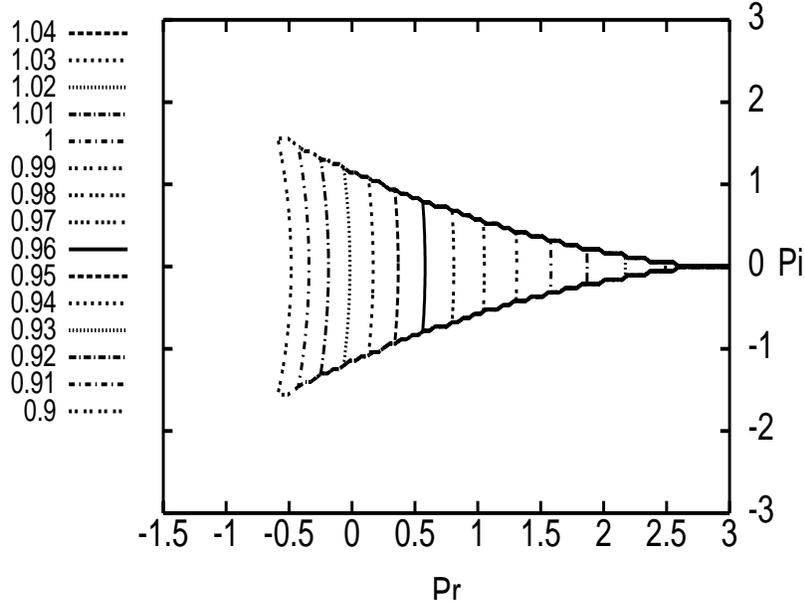,height=8cm,width=12cm,angle=-90,clip=}
\end{center}
\vspace*{-0.5truecm}
\caption{Isocontour lines for $1/V \log |\det|$ in the toy model.}
\end{figure}

\begin{figure}[!t]
\vspace*{-0.5truecm}
\begin{center}
\epsfig{bbllx=3.75truecm,bblly=8.8truecm,bburx=16.5truecm,bbury=21.5truecm,
file=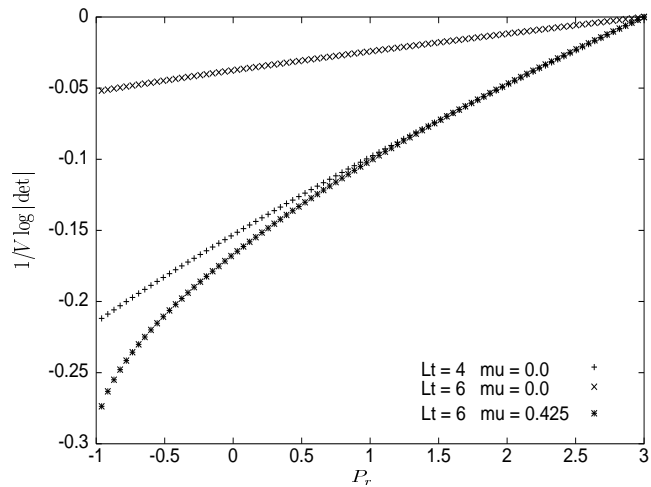,height=7cm,width=9cm,clip=}
\end{center}
\vspace*{-1.0truecm}
\caption{Logarithm of the modulus of the determinant for $P_i=0$ in the toy
model: a decrease in temperature (from middle to top curve) can be nicely 
compensated by an increase in the chemical potential (bottom curve).}
\end{figure}
\begin{figure}[!h]
\begin{center}
\epsfig{bbllx=5.75truecm,bblly=7.68truecm,bburx=16.8truecm,bbury=22.95truecm,
file=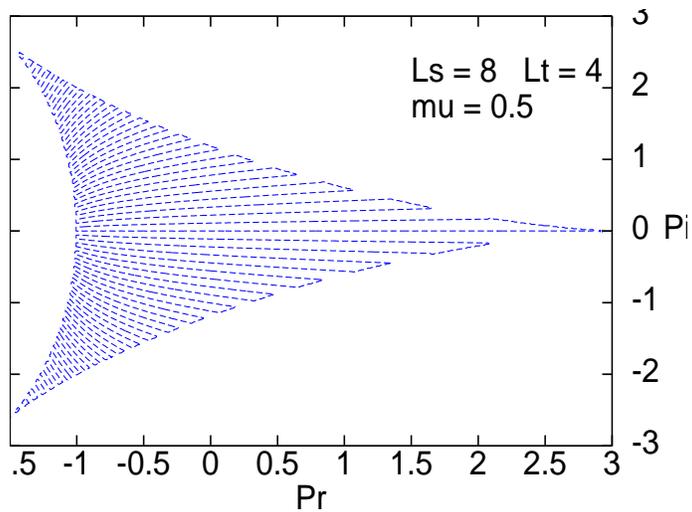,height=7cm,width=9cm,angle=-90,clip=}
\end{center}
\vspace{-0.75truecm}
\caption{Lines where $\phi(\det))=2k\pi$ in the $(P_r,P_i)$ plane for the 
toy model, on an $8^3\times4$ lattice. The phase is proportional to the lattice
volume, so that the lines depicted become denser as the volume increases. }
\end{figure}

Fig.3 shows $\frac{1}{V} \log |\det|$, as a function of the Polyakov loop.
It is maximum in the free field case $A_0=0, P=(3,0)$,
in accord with the theorem of E. Seiler \cite{Seiler}.
Note that isocontour lines are almost vertical and almost equally spaced, 
which implies that $\log |\det|$ is almost a linear function of $P_r$ alone:
$\frac{1}{V} \log |\det| \sim c(\mu,T,m_q) P_r + c_0$.
What is particularly 
interesting is the variation of the slope $c(\mu,T,m_q)$ with
$\mu$ and $T$, since this quantity characterizes the amount by which
larger Polyakov loop values $P_r$ are favored, i.e. the ordering effect
of the fermion determinant. Fig.4 shows clearly that a similar ordering 
effect can be achieved by increasing the temperature
{\em or} the chemical potential.

The oscillatory behavior of the determinant is displayed in Fig.5, which
shows the phase $\frac{1}{V} \phi(\det)$, as a function of the Polyakov loop.
The isocontour lines correspond to increases of the phase by $2 \pi$.
They are almost parallel
and equally spaced, confirming $\phi(\det) \propto P_i$ in this model.
This figure makes it clear how the sign problem occurs: as $V$ increases, the
quenched measure $\exp(-S_g)$ allows for fluctuations in $P_i$ of 
${\cal O}(1/\sqrt{V})$, which are sufficient to rotate the phase $\phi(\det)$
by ${\cal O}(\sqrt{V})$, and drive $\langle \cos(\phi(\det)) \rangle$ to zero.

\begin{figure}[!b]
\begin{center}
\epsfig{bbllx=3.5truecm,bblly=9.8truecm,bburx=16.5truecm,bbury=21.5truecm,
file=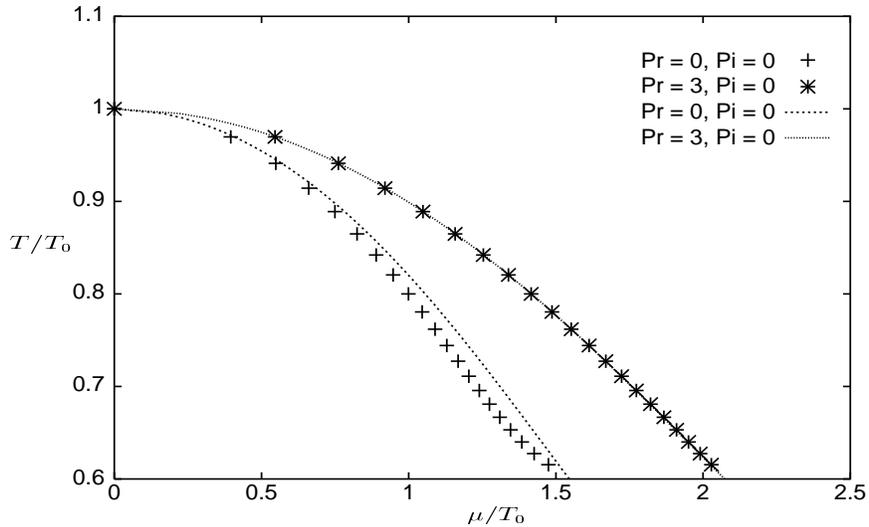,height=7cm,width=12cm,clip=}
\caption{Heuristic phase boundary obtained from the toy model.
The lines are derived from the continuum calculation of \cite{Pisarski},
the data points are obtained from the lattice. The 2 boundaries correspond
to different critical values of $P_r$.}
\end{center}
\end{figure}

Assume now that the $\mu=0$ QCD theory admits a deconfinement transition
at $T=T_0$, characterized by 2 degenerate minima of the free energy at
neighboring (or degenerate) real values $P_1 \leq P_2$ of the Polyakov loop.
If the chemical potential $\mu$ is turned on, the fermionic determinant will
favor the minimum at $P_2$ because of the ordering effect of $\mu$ (Fig.4).
The balance between $P_1$ and $P_2$ can be restored by a compensating decrease
in temperature. This is achieved when the derivative 
$\frac{d}{dP_r} log(|\det(\mu,T)|)|_{P_1}$ keeps its $(\mu=0,T=T_0)$ value. 
In this way, a heuristic phase boundary $\mu_c(T)$ can be obtained, as
in Fig.6. The phase boundary can be obtained by the same arguments
using the continuum expression for the fermion determinant in a gauge
field background $A_0$ (Eq. [12] of Ref. \cite{Pisarski}).
The lines of Fig.6 come out from the continuum expression,
and the discrete points from the lattice expression for
staggered fermions. The two sets of points correspond to $P_1=3$ and $P_1=0$,
to give some measure of our systematic uncertainties. The phase diagram
shows a quadratic dependence of the critical temperature versus
$\mu$. It even is quantitatively plausible. 
Nonetheless, several crude approximations have been made: 
all fluctuations of the gauge field have been ignored except the zero-mode
of the Polyakov loop;
and the variation of the quenched measure with $T$ is neglected.
The failure of this crude approximation becomes obvious as the temperature
is lowered: the critical chemical potential $\mu_c(T)$ diverges as 
$T \rightarrow 0$ \footnote{This is brought about by the breakdown of the 
approximation for the fermion determinant, by which
$\log(|\det(\mu=0)|)$ varies as
$T^3$ (see \cite{Pisarski}) instead of
$1/T$ as it should as $T \rightarrow 0$.}. Therefore, it is likely that
$\mu_c(T)$ consistently overestimates the true critical $\mu$.

\begin{figure}[!t]
\begin{center}
\epsfig{bbllx=3.0truecm,bblly=8.8truecm,bburx=16.5truecm,bbury=21.5truecm,
file=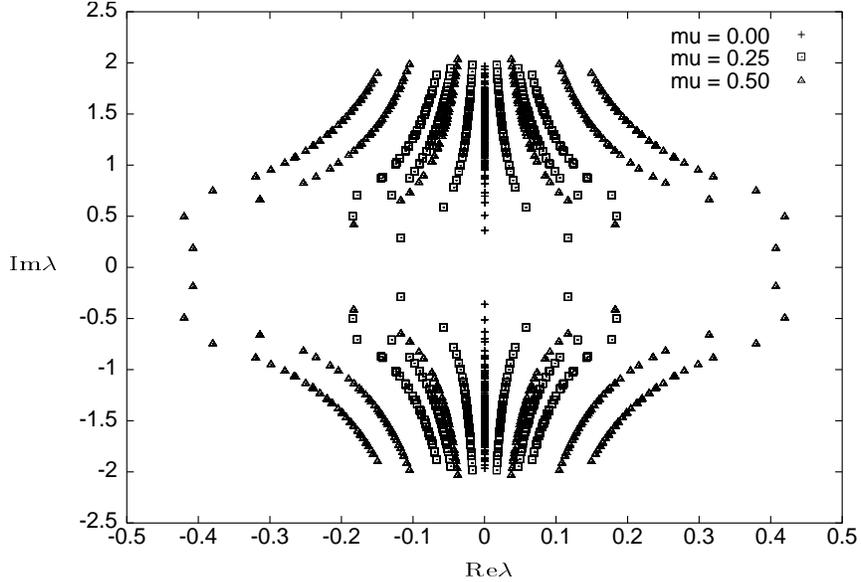,height=8cm,width=12cm,clip=}
\caption{Spectrum of the fermion matrix for $(P_r,P_i) = (0,0)$ on a
$8^3\times4$ lattice, for three values of the chemical potential $\mu$.
Eigenvalues are repelled from the origin as $\mu$ increases, indicating
chiral symmetry restoration.}
\end{center}
\end{figure}

However, our toy model may still provide some qualitative insight as to 
what happens at low temperatures. As $T \rightarrow 0$, the quenched measure
for the Polyakov loop becomes Gaussian around the origin:
$Prob(P) \propto \exp(-c V (P_r^2 + P_i^2))$. 
On the other hand, note the slight curvature of the contour lines in Fig.3.
They indicate that, if $P_r$ fluctuates little around 0, $log(|\det|)$ 
{\em increases} with $|P_i|$: the determinant favors large values of $P_i$,
whereas the quenched measure favors small values. The partition function
has the form
\be
\int dP_i ~e^{-c V P_i^2} ~~ e^{i V P_i b(\mu) + V d(\mu) P_i^2}
\label{ansatz}
\ee
with the first factor coming from the quenched measure, the second one from
the fermion determinant. If for some value of $\mu$, $d(\mu) = c$, then
a phase transition ensues. The major contribution to the partition function,
instead of coming from the region $P_i \sim 0$, will come from the large
values of $P_i$. To support this plausible scenario, Fig.7 shows the Dirac
spectrum in the complex plane for our toy model, at $(P_r,P_i) = (0,0)$,
for increasing values of $\mu$. As $\mu$ increases, eigenvalues 
are repelled from the origin, which would be consistent with $\psibarpsi = 0$
and chiral symmetry restoration at large $\mu$.

Finally, a remarkable feature of our heuristic phase diagram is that its
form remains unchanged for $SU(2)$. There is no difference
at all if one chooses for $P_1$ its free field value (upper curve in Fig. 6).
For any other value the difference between the phase boundary for the
$SU(2)$ and $SU(3)$ cases is a constant factor. 
An exact $SU(2)$ phase diagram should 
soon be available for comparison, since $SU(2)$ simulations do not suffer from
the sign problem.

\section{Constrained Monte Carlo}

To go beyond our toy model, in $SU(3)$, one must address the sign problem.
We have considered the introduction of a constraint on $P_i$ in the partition
function. Namely we want to study by Monte Carlo the modified partition
function
\be
Z' = \int [dU] ~e^{-S_g(U)} ~{\rm det}^2(U,\mu) ~\delta(P_i(U))
\ee
The benefits are clear: as seen in Section I, the fluctuations in 
$\phi(\det)$ will be reduced, especially for heavy quarks. 
In the same way that the microcanonical and canonical partition functions
become equivalent in the thermodynamic limit, the constraint on $P_i$
becomes irrelevant as $V \rightarrow \infty$, provided $P_i$
is frozen to its mean value given by the saddle point
of its effective potential. 
However, at finite density, the mean value of $P_i$ is 
imaginary, as can be seen from Eq.(\ref{ansatz}). Therefore, our constraint 
introduces some systematic error. This can be traced to the following.

\begin{figure}[!t]
\begin{center}
\epsfig{bbllx=4.0truecm,bblly=9.8truecm,bburx=16.5truecm,bbury=21.5truecm,
file=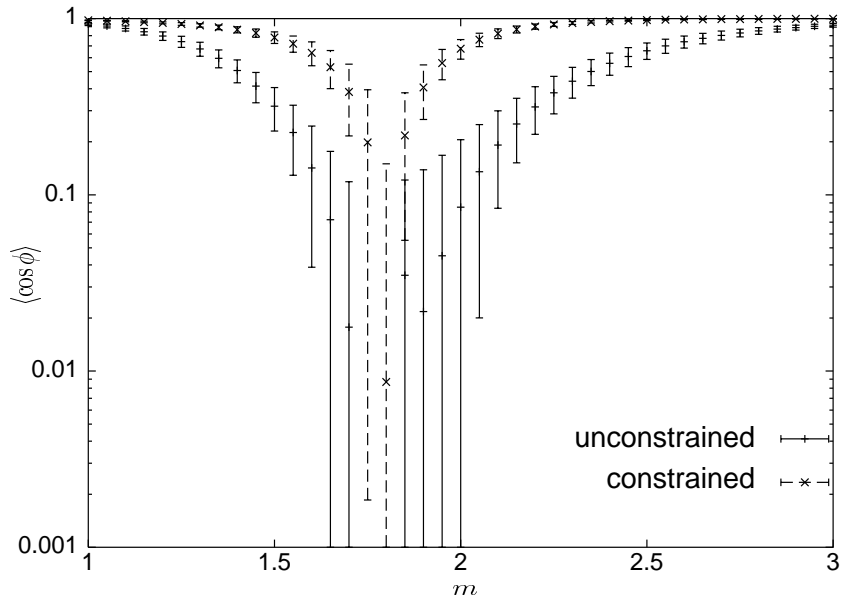,height=8cm,width=12cm,clip=}
\caption{Average of $\cos(\phi(\det))$ as a function of the fermion mass
$m$ in two quenched simulations on a $4^3\times 6$ lattice at $\beta=1$ and
$\mu=1.5$. Both simulations have equal statistics. 
One of them includes the constraint $P_i=0$.}
\end{center}
\end{figure}

The free energy $F_q$ of a single quark at finite $\mu$ is given by the 
Polyakov
loop:
\be
e^{-\beta F_q} = \langle {\rm Tr} P \rangle = P_r + i P_i
\ee
where $P_i$ is pure imaginary. On the other hand, for an antiquark one has
\be
e^{-\beta F_{\bar q}} = \langle {\rm Tr} P^\dagger \rangle = P_r - i P_i
\ee
Constraining $P_i=0$ thus enforces equality of the 2 free energies. This is
a good approximation in 2 regimes: \\
- $|P_r| \ll |P_i|$, i.e. high temperature or small chemical potential; \\
- $|P_r| = |P_i| = 0$, i.e. zero temperature, for any $\mu < \mu_c$. \\
This second regime is particularly interesting and difficult to investigate.

To test our ideas, we have implemented a constrained Hybrid Monte Carlo
algorithm, and performed quenched $SU(3)$ simulations, measuring the average
``sign'' $\langle \cos(\phi(\det)) \rangle$ with and without the
constraint $P_i = 0$. 
Fig.8 shows our results as a function of the quark mass, for chemical potential
$\mu = 1.5$ in lattice units, at $T \sim 0$. The relative error on
$\langle \cos(\phi(\det)) \rangle$, which propagates to all observables, is 
reduced by over an order of magnitude as one approaches the phase transition.

\section{Conclusion} 

We have exhibited the strong correlation present in static QCD between the
phase of the fermion determinant and the imaginary part of the Polyakov loop.
This correlation motivated us to study a toy model which keeps the zero-mode
of the Polyakov loop as the only degree of freedom. This toy model clearly
shows the ordering effect of the chemical potential and the oscillations of
the fermionic measure. A heuristic phase diagram can be obtained, and a 
mechanism for a $T=0$ phase transition suggests itself.
Finally, constrained Monte Carlo simulations where the imaginary part of the
Polyakov loop is fixed to zero provide an important reduction of the sign
problem, and allow to go beyond static QCD.

We thank the authors of Ref.\cite{Blum}, especially Doug Toussaint, for
making their data available to us. Ph. de F. thanks Mike Creutz for helpful
discussions.


\end{document}